\documentclass[final,3p,times,twocolumn]{elsarticle}

\biboptions{numbers,sort&compress}

\usepackage[linesnumbered,ruled]{algorithm2e}

\pdfoutput=1
\usepackage[table,xcdraw]{xcolor}
\usepackage{graphicx}
\usepackage{epstopdf}
\usepackage{mathrsfs}
\usepackage{amssymb}
\usepackage{verbatim}
\usepackage{color}
\usepackage{hhline}
\usepackage{stackrel}
\usepackage[pdfencoding=auto]{hyperref}
\usepackage{array,multirow}
\usepackage{colortbl}
\usepackage{hhline}
\usepackage{lipsum}
\usepackage{amsmath}
\usepackage{multirow}

\DeclareMathOperator*{\argmin}{arg\,min}

\newcommand{\f}{\mathbf{f}}

\newcommand{\beq}{\begin{equation}}
\newcommand{\eeq}{\end{equation}}
\newcommand{\ga}{\lower.7ex\hbox{$\;\stackrel{\textstyle>}{\sim}\;$}}
\newcommand{\la}{\lower.7ex\hbox{$\;\stackrel{\textstyle<}{\sim}\;$}}

\usepackage{hyperref}
\usepackage{amsmath,bm}
\usepackage{physics}
\usepackage{tikz}
\allowdisplaybreaks

\hypersetup{
    colorlinks = true,
    citecolor = {blue},
    linkcolor = {blue},
    urlcolor = {blue},
}

\usetikzlibrary{shapes.geometric, arrows}

\tikzstyle{genmod} = [rectangle, rounded corners, 
minimum width=2.5cm, 
minimum height=2.5cm, text centered, 
draw=black, fill=blue!40]

\tikzstyle{oracle} = [rectangle, rounded corners, 
minimum width=2.5cm, 
minimum height=2.5cm, text centered, 
draw=black, fill=red!40]

\tikzstyle{loss} = [rectangle, rounded corners, 
minimum width=2.5cm, 
minimum height=3.5cm, text centered, 
draw=black, fill=green!40]

\tikzstyle{circlenode} = [circle,  
minimum width=0.01cm, 
minimum height=0.01cm, text centered, 
draw=black, fill=black]

\tikzstyle{circlenode2} = [rectangle,  
minimum width=0.01cm, 
minimum height=0.01cm, text centered, 
draw=none, fill=none]

\usepackage{etoolbox}

\makeatletter
\patchcmd{\@classz}
  {\CT@row@color}
  {\oldCT@column@color}
  {}
  {}
\patchcmd{\@classz}
  {\CT@column@color}
  {\CT@row@color}
  {}
  {}
\patchcmd{\@classz}
  {\oldCT@column@color}
  {\CT@column@color}
  {}
  {}
\makeatother

\usepackage{colortbl}

\makeatletter

    \def\CT@@do@color{%
      \global\let\CT@do@color\relax
            \@tempdima\wd\z@
            \advance\@tempdima\@tempdimb
            \advance\@tempdima\@tempdimc
    \advance\@tempdimb\tabcolsep
    \advance\@tempdimc\tabcolsep
    \advance\@tempdima2\tabcolsep
            \kern-\@tempdimb
            \leaders\vrule
                    \hskip\@tempdima\@plus  1fill
            \kern-\@tempdimc
            \hskip-\wd\z@ \@plus -1fill }
    \makeatother

\journal{Physics Letters B}

\begin{document}

 \SetKwComment{Comment}{/* }{ */}
 \SetKw{KwFrom}{from}
 \SetKw{KwTerminate}{terminate}
 \SetKw{KwAppend}{append} 
 \SetKw{KwContinue}{continue}
 \SetKw{KwClear}{clear} 
 \SetKw{KwInput}{Input} 
 \SetKw{KwParameters}{Parameters}  
 \SetKw{KwGoto}{goto} 
 \SetKw{KwStop}{stop} 
 \SetKw{KwWhile}{while(true)}  
\begin{frontmatter}

\title{Accelerated Discovery of Machine-Learned Symmetries:\texorpdfstring{\\}{} Deriving the Exceptional Lie Groups \texorpdfstring{$G_2$}{G2}, \texorpdfstring{$F_4$}{F4} and \texorpdfstring{$E_6$}{E6} }

\author[UF]{Roy T.~Forestano\fnref{contribution}} 
\author[UF]{Konstantin T.~Matchev\fnref{contribution}}
\author[UF]{Katia Matcheva\fnref{contribution}}
\author[UF]{Alexander Roman\fnref{contribution}}
\author[UF]{Eyup B.~Unlu\fnref{contribution}}
\author[UF]{Sarunas~Verner\fnref{contribution}}

\fntext[contribution]{All authors share equal contributions to this paper.}
\affiliation[UF]{organization={Institute for Fundamental Theory, Physics Department, University of Florida},
            city={Gainesville},
            state={FL},
            postcode={32611}, 
            country={USA}}

\begin{abstract}
Recent work has applied supervised deep learning to derive continuous symmetry transformations that preserve the data labels and to obtain the corresponding algebras of symmetry generators. This letter introduces two improved algorithms that significantly speed up the discovery of these symmetry transformations. The new methods are demonstrated by deriving the complete set of generators for the unitary groups $U(n)$ and the exceptional Lie groups $G_2$, $F_4$, and $E_6$. A third post-processing algorithm renders the found generators in sparse form. We benchmark the performance improvement of the new algorithms relative to the standard approach. Given the significant complexity of the exceptional Lie groups, our results demonstrate that this machine-learning method for discovering symmetries is completely general and can be applied to a wide variety of labeled datasets.
\end{abstract}




\end{frontmatter}

\section{Introduction}

The beginning of the 20th century significantly changed theoretical physics. It became evident that the laws of nature are closely tied to principles of symmetry~\cite{Gross1996}. Emmy Noether taught us that a continuous symmetry within any physical system inherently implies a conservation law, providing profound insights about the system~\cite{Noether1918}. This concept has deepened our understanding of fundamental physics, from the simplest principles of classical mechanics to the intricacies of the Universe at large. In particle physics, these symmetries serve as the foundation that organizes the particles and their associated interactions. They are instrumental in guiding theoretical physicists in exploring possible extensions of the Standard Model~\cite{Peskin:1997ez,Csaki:2018muy}.

Group theory has traditionally served as the framework for studying symmetries \cite{wigner1959group}. Among the numerous types of classical Lie groups, the special orthogonal groups $SO(n)$ and the special unitary groups $SU(n)$ are most commonly used in particle physics. While some classical Lie groups describe symmetries observed in nature, the exceptional Lie groups \cite{Ramond:1976aw} open up new possibilities for theoretical physics, potentially enabling us to describe new kinds of gauge theories, including grand unified theories (GUTs) \cite{Slansky:1981yr}. 

There exist five exceptional Lie groups --- $G_2$, $F_4$, $E_6$, $E_7$, and $E_8$, all of which have found various applications in theoretical physics \cite{ramond2003exceptional}. For example, the smallest among them, $G_2$, has been widely used in string theory, e.g., in the studies of seven-dimensional manifolds known as $G_2$-manifolds, used in M-theory compactifications~\cite{Acharya:1998pm, Acharya:2004qe, Atiyah:2001qf, Halverson:2015vta}. The next exceptional group, $F_4$, plays an important role in Jordan algebra theory~\cite{Catto:2013xba} and has also been used in the context of gauge theories~\cite{Shahlaei:2018lth, Rafibakhsh:2017zfm}. The exceptional Lie group $E_6$ has been broadly employed in GUTs~\cite{Gursey:1975ki, Croon:2019kpe}. The two remaining exceptional groups, $E_7$ and $E_8$, have also been considered as potential GUT candidates~\cite{Gursey:1976dn, Bars:1980mb}.

The use of machine learning (ML) to uncover and recognize symmetries in datasets has recently sparked considerable interest~\cite{Iten1807, Krippendorf:2020gny, Liu:2020omw,Barenboim:2021vzh,Dillon:2021gag,Liu:2021azq,Desai:2021wbb,Craven:2021ems,Moskalev2210.04345,Forestano:2023fpj,Roman:2023ypv,Forestano:2023qcy}. Specific applications to group theory include calculating tensor products and branching rules of irreducible representations of Lie groups \cite{Chen:2020jjw} or testing for the presence of a conjectured Lie group symmetry in the data \cite{Liu:2021azq,Craven:2021ems}. More recent work has focused on discovering from first principles the Lie group generators reflecting symmetries in the data \cite{Moskalev2210.04345,Forestano:2023fpj,Roman:2023ypv,Forestano:2023qcy}. In this Letter, we introduce new algorithms that significantly enhance the symmetry discovery process outlined in~\cite{Moskalev2210.04345,Forestano:2023fpj, Roman:2023ypv,Forestano:2023qcy}. After introducing the problem in Section~\ref{sec:setup}, in Section~\ref{sec:tricks} we describe the new methods and demonstrate their advantages on the unitary groups $U(n)$. Then in Sections~\ref{sec:G2}, \ref{sec:F4} and \ref{sec:E6} we consecutively consider three of the five exceptional Lie groups: $G_2$, $F_4$, and $E_6$. We derive the complete set of generators, in sparse form, which preserve the respective polynomial invariants. Our approach is completely general, and could be easily extended to the remaining two exceptional Lie groups $E_7$ and $E_8$. Section~\ref{sec:summary} is reserved for our summary.

\section{Problem Description}
\label{sec:setup}

The classical groups are the linear groups of transformations over the real numbers $\mathbb R$, the complex numbers $\mathbb C$, and quaternions $\mathbb  H$. Consequently, a symmetry transformation operates on a feature vector ${\mathbf x}\equiv \{x^{(1)}, x^{(2)},\ldots, x^{(n)}  \}$, where ${\mathbf x}\in \mathbb R^n$ (${\mathbf x}\in \mathbb C^n$) for real (complex) representations. To encapsulate the effect of a group transformation on ${\mathbb R}^n$ or ${\mathbb C}^n$, we examine a representative set of $m$ points $\left\{\mathbf{x}\right\} \equiv \left\{\mathbf{x}_1,\mathbf{x}_2,\ldots,\mathbf{x}_m\right\}$ sampled from a finite domain. The selection of a sampling distribution, as well as the size and location of the domain, are inconsequential. We adopt a standard normal distribution and pick $m$ on the order of several hundred, depending on the complexity of the problem.

The classical groups can be defined in terms of polynomial invariants over their respective fields. For example, the orthogonal group $O(n)$ preserves the values of the polynomial oracle,
\beq
\varphi_{O}(\mathbf{x}) 
\; \equiv \; |\mathbf{x}|^2 = \sum_{j=1}^n \bigl(x^{(j)}\bigr)^2, 
\quad 
x^{(j)}\in \mathbb R \,,
\label{oracle:O}
\eeq
the Lorentz group in $n=4$ dimensions preserves 
\beq
\varphi_{L}(\mathbf{x}) 
\equiv 
\bigl(x^{(1)}\bigr)^2 
-\bigl(x^{(2)}\bigr)^2 
-\bigl(x^{(3)}\bigr)^2 
-\bigl(x^{(4)}\bigr)^2,
~~
x^{(j)}\in \mathbb R \, ,
\label{oracle:L}
\eeq
and the unitary group $U(n)$ preserves
\beq
\varphi_{U}(\mathbf{x}) 
\; \equiv \; \sum_{j=1}^n \bigl(x^{(j)}\bigl)^\ast x^{(j)}, 
\quad 
x^{(j)}\in \mathbb C \, .
\label{oracle:U}
\eeq

A symmetry transformation $\f$ is a map $\mathbf{x}' = \f(\mathbf{x})$ that preserves the respective oracle 
(\ref{oracle:O}-\ref{oracle:U}) everywhere, or in our case, for each of the sampled $m$ points:
\beq
\varphi(\mathbf{x}'_i) \; \equiv \; \varphi(\f(\mathbf{x}_i)) = \varphi(\mathbf{x}_i), \quad \forall i \,= \,1,2,\ldots,m \, .
\eeq
The basic task is to find such a symmetry map $\f$ in parametric form. To focus on the {\em generators} of the symmetry transformation group, $\f$ is linearized by considering infinitesimal transformations $\delta{\f}$ in the vicinity of the identity transformation $\mathbb{I}$:
\beq
\delta{\f} \; \equiv \; \mathbb{I} + \varepsilon \, {\mathbb G} \, ,
\label{eq:deltaf}
\eeq
where $\varepsilon$ is an infinitesimal parameter and ${\mathbb G}$ is an $n\times n$ matrix. After training with the invariance loss function $L_\text{inv}({\mathbb G} , \{\mathbf x\})$ defined in (\ref{eq:LossInvariance}), the components of ${\mathbb G}$ evolve towards their {\em trained} values, thereby producing a valid symmetry generator~\cite{Forestano:2023fpj}
\beq
\mathbb{J} \; \equiv \;  \argmin_{\mathbb G}
\Bigl(L({\mathbb G}, \{\mathbf x\}) \Bigr).
\label{eq:trainingcondition}
\eeq

\begin{figure}[t]
    \centering
    \includegraphics[width=0.90\columnwidth]{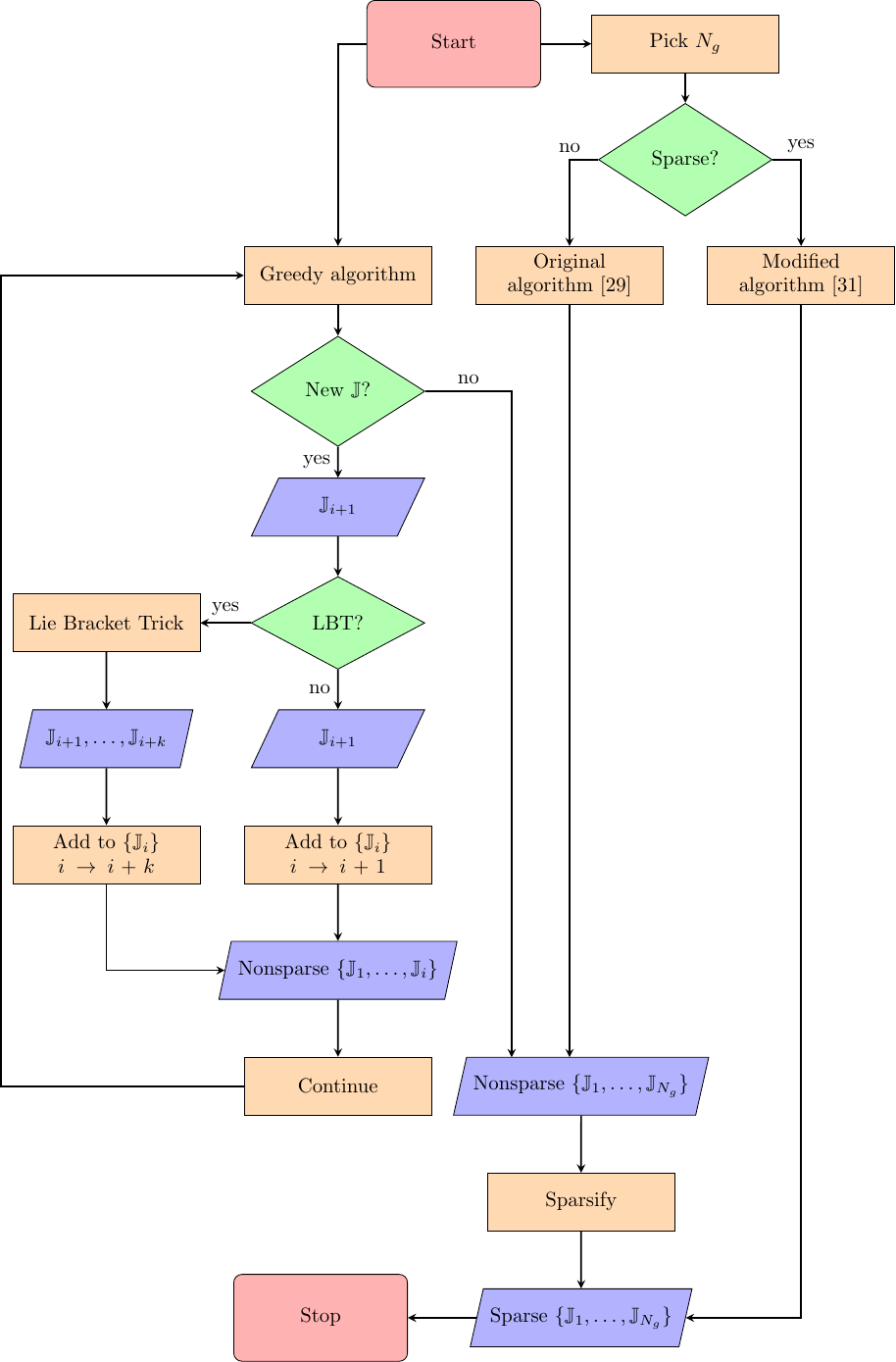}
    \caption{Flowchart illustrating the algorithms discussed in Section~\ref{sec:tricks} (left side) and the algorithms from Refs.~\cite{Forestano:2023fpj} and \cite{Forestano:2023qcy} (right side).}
    \label{fig:flowchart}
\end{figure}

As shown in the flowchart of Figure~\ref{fig:flowchart}, this idea was used in \cite{Forestano:2023fpj} to simultaneously train $N_g$ generators, which were designed to be normalized, orthogonal to each other, and to form a closed algebra, by adding suitable terms to the loss function. However, as indicated in Figure~\ref{fig:flowchart}, the resulting set of $N_g$ symmetry generators $\{\mathbb J_\alpha\}$ was not sparse. An improvement was implemented in \cite{Forestano:2023qcy}, by adding an extra term to the loss function, Eq.~(\ref{eq:LossSparsity}), which encourages the learning of sparse generators in the canonical textbook form.

\section{Iterative Constructions of Symmetry Generators}
\label{sec:tricks}

The advantage of learning $N_g$ symmetry generators in one go is that one can impose the closure condition in the loss function and thus guarantee that the learned set $\{\mathbb J_\alpha\}$ is closed. However, there are drawbacks to this approach as well. First, due to the large number of trainable parameters ($\sim n^2\times N_g$), the training is challenging and becomes prohibitively slow for large groups (large $N_g$) and/or high dimensions (large $n$). For this reason, in Section~\ref{sec:greedy} we explore an alternative ``greedy'' approach, which trains one generator at a time. Secondly, a closed algebra of generators only exists for certain values of $N_g$, which are a priori unknown, and with the previous algorithms, would have to be guessed by trial and error. In contrast, the greedy approach automatically finds the largest possible closed set of generators. Furthermore, we can use the closure condition to find additional generators directly, without any training. We refer to this last method as the ``Lie bracket trick" (LBT), which is described in Section~\ref{sec:LBT}. In principle, both of the two new methods could accommodate the sparsity condition (\ref{eq:LossSparsity}). However, we find that the training is more efficient when we look for nonsparse generators, therefore we postpone their sparsification to a postprocessing step which we describe in Section~\ref{sec:sparsification}.

\begin{algorithm}[t]
\caption{The greedy algorithm. }
\label{alg:greedy}
\SetAlgoLined
$\KwParameters$: $\lambda, L_{min}, N_{epochs}$\;
$\{\mathbb J\}$ $\gets$ $[]$\;
$\mathcal W \gets \mathcal W_{initial}\sim {\mathcal N}$\;
\For{i $\KwFrom$ 1 $\KwTo$ $N_{epochs}$}{

    L $\gets L_{\text{greedy}}(\mathbb G({ \mathcal W}),\{\mathbb J\},\mathbf{x})$ \;        
    \If{$L<L_{min}$}{
            $\KwAppend$ $\mathbb G(\mathcal W)$ $\KwTo$ $\{\mathbb J\}$\;
            \KwGoto 3;
            }
    $\mathcal W$ $\gets \mathcal W - \lambda \nabla_{\mathcal W}L_{\text{greedy}}$\;            
}      
\KwStop            
\end{algorithm}

\subsection{Greedy Algorithm}
\label{sec:greedy}

The basic idea of the greedy algorithm is illustrated in Figure~\ref{fig:flowchart} and the corresponding pseudocode is shown in Algorithm~\ref{alg:greedy}. We use Eq.~(\ref{eq:trainingcondition}) to train one generator $\mathbb G(\mathcal W)$ at a time, where $\mathcal W$ denotes the trainable parameters of the matrix $\mathbb G$. The generator $\mathbb G$ is required to i) preserve the oracle, ii) be normalized, and iii) be orthogonal to the set of generators $\{\mathbb J_\alpha\}$ already found so far. Therefore, the loss function is
\beq 
L_\text{greedy} (\mathbb G(\mathcal W), \{\mathbb J\}, \{\mathbf x\}) = 
L_\text{inv} + L_\text{norm} + L_\text{ortho},
\label{eq:greedyloss} 
\eeq
where $L_\text{inv}$, $L_\text{norm}$ and $L_\text{ortho}$ are defined in (\ref{eq:LossInvariance}), (\ref{eq:LossNormalization}) and (\ref{eq:LossOrthogonality}), respectively.

As shown in Figure~\ref{fig:flowchart}, at the start of the algorithm the learned set $\{\mathbb J\}$ is empty ($i=0$). At this point, the orthogonality loss $L_\text{ortho}$ is not applicable and is turned off. The minimization in Eq.~(\ref{eq:trainingcondition}) will yield a single new symmetry generator $\mathbb J_\text{new}=\mathbb J_1$, the first to be added to the learned set $\{\mathbb J\}$. Now the orthogonality loss $L_\text{ortho}$ is turned on and the process continues until the algorithm fails to find a new valid symmetry generator $\mathbb J_\text{new}$. The resulting set $\{\mathbb J_1,\ldots,\mathbb J_{N_g}\}$ is the largest orthogonal set of nontrivial symmetry generators. For the orthogonal groups $O(n)$ with the oracle~(\ref{oracle:O}), this procedure terminates when $N_g = n(n-1)/2$, while for the unitary groups $U(n)$ with the oracle~(\ref{oracle:U}), it stops at $N_g = n^2$.

\subsection{Lie Bracket Trick}
\label{sec:LBT}

Algorithm~\ref{alg:lie} shows the pseudocode for the Lie bracket trick algorithm, which leverages the existing group structure among the symmetry generators. As shown with the left side branch in Figure~\ref{fig:flowchart}, the LBT can be (optionally) applied in conjunction with the greedy algorithm, once two orthogonal generators, $\mathbb J_1$ and $\mathbb J_2$, have been found. At that point, one can compute their commutator (Lie bracket) $\left[\mathbb J_1,\mathbb J_2\right]$, with three possible outcomes. First, if the commutator is zero, we gain nothing from the LBT and must return to the greedy algorithm. If, however, the chosen pair does not commute, the result may include a component outside the span of the generators discovered so far, which would lead to a new valid symmetry generator that we can extract for free. To isolate this component, we employ Gram-Schmidt orthogonalization (line 11 in Algorithm~\ref{alg:lie}). If the result is zero, we are again out of luck and must return to the greedy method. However, if the out-of-span component is nonzero, it becomes (after the normalization in line 13) a new generator which can be added to the set $\{J\}$. This process can be repeated for every pair of known generators, including the newly found ones via the LBT method. The LBT algorithm terminates when all possible Lie brackets close in the current set $\{\mathbb J\}$. We note that when using the two algorithms~\ref{alg:greedy} and \ref{alg:lie} together, one should choose judiciously the respective values of the loss thresholds $L_{min}$.

\begin{algorithm}[t]
\caption{\mbox{The Lie bracket trick (LBT) algorithm.}\label{alg:lie}}
\SetAlgoLined
$\KwInput$: $\{J_1, \ldots, J_i\}$: known algebra; $J_{i+1}$: new generator\;
$\KwAppend$ $J_{i+1}$ $\KwTo$  $G$\;
\Repeat{$|G|=0$}{
$k \gets |G|$\;
$i \gets |J|$\;
$\KwAppend$ G $\KwTo$  J\;
$\KwClear$ G\;
\For{p $\KwFrom$ 1 $\KwTo$ i}{
    \For{q $\KwFrom$ i+1 $\KwTo$ i+k}{
        C $\gets J_pJ_q - J_qJ_p$\;
        C $\gets \text{C}- \sum_{g\in {J}} \frac{g}{||g||} \times(\text{C} \cdot  g)$\;
        \If{$||C||\ne 0$}
        {
        C $\gets \frac{\text{C}}{||\text{C}||}$\;
        \If{$L_{inv}(C,{\bf{x}})<L_{min}$}{
            $\KwAppend$ C $\KwTo$  G\;
}}}}}
\end{algorithm}

\subsection{Sparsification}
\label{sec:sparsification}

Suppose we have already found a set of $N_g$ generators $J_{\alpha}$, $\alpha=1,2,\ldots,N_g$. We can transform them to a new sparse basis, $\tilde J_\alpha$, by rotating with an orthogonal $N_g\times N_g$ matrix $O$,
\begin{equation}
    \tilde J_\alpha (O)= O_{\alpha\beta}J_{\beta}.
    \label{eq:sparse_rotation}
\end{equation}
In analogy to (\ref{eq:LossSparsity}), the loss function for the sparsification of the generators can be defined as
\begin{align}
    L_\text{sp} (O)= \sum_{\alpha =1}^{N_g} \sum_{\substack{j,j'=1\\k,k'=1}}^n \left|\tilde{J}_\alpha^{(jk)} (O)\tilde{J}_\alpha^{(j'k')}(O)\right| \left(1-\delta_{jj'}\delta_{kk'} \right).
    \label{eq:sparsity_loss}
\end{align}
It takes into account all possible pairs of entries in each transformed generator $\tilde{J}_\alpha$. We minimize over this loss to obtain the desired orthogonal transformation $O$, and subsequently, apply this transformation to the original generators $J_\alpha$ to find the sparse generators $\tilde{J}_\alpha$ as in (\ref{eq:sparse_rotation}).

\subsection{Timing tests}

\begin{figure}[t]
    \centering
    \includegraphics[width=0.95\columnwidth]{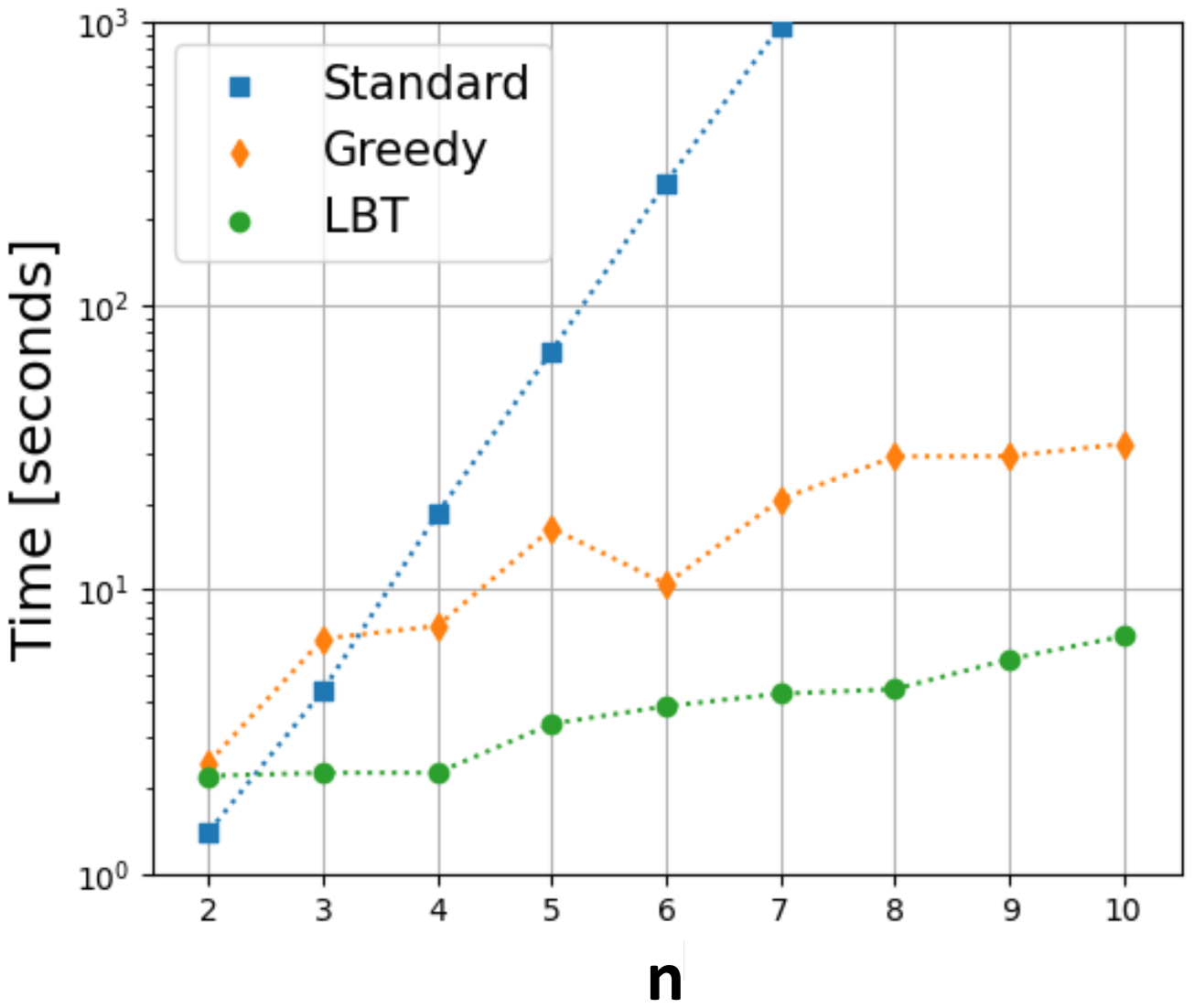}
    \caption{Comparison of the running times of the different algorithms in finding the full algebras of the $U(n)$ family: the standard algorithm \cite{Forestano:2023qcy} (blue squares), the greedy algorithm \ref{alg:greedy} (orange diamonds) and the Lie bracket trick algorithm \ref{alg:lie} (green  circles).
    }
    \label{fig:timing}
\end{figure}

The main advantage of the two new algorithms is that they significantly speed up the training procedure. To quantify this improvement, in Figure~\ref{fig:timing} we present the results from timing tests on a personal laptop for different $U(n)$ groups, using the three approaches discussed earlier: the standard algorithm \cite{Forestano:2023qcy} (blue squares), the greedy algorithm \ref{alg:greedy} (orange diamonds) and the LBT algorithm \ref{alg:lie} (green  circles). The plot shows the time in seconds that it took to learn all $n^2$ generators for the $U(n)$ group, as a function of the dimension $n$, at a learning rate  $2.5\times 10^{-2}$. The generators were found in sparse form by applying the post-processing step from Section~\ref{sec:sparsification}, whose duration was included in the total time shown in Figure~\ref{fig:timing}. We observe that for small $n$ the performance of all three methods is comparable, but for large $n$ the new methods offer significant improvement. In particular, for $n\sim 10$ the standard method would require training for days, while the new methods reduce the training time to less than a minute.

\section{The Exceptional Group \texorpdfstring{$G_2$}{G2}}
\label{sec:G2}

The ML approach described in the previous two sections can be used to discover the orthogonal groups $O(n)$ \cite{Forestano:2023fpj} and the unitary groups $U(n)$ \cite{Forestano:2023qcy}. The method can also be generalized to the case of vector (i.e., multicomponent) oracles \cite{Roman:2023ypv}. We shall now apply it to exceptional (non-classical) Lie algebras, which have relatively large number of generators, and would benefit from the speed-up offered by the new algorithms. We consider three of the five exceptional Lie groups: $G_2$ in this section, $F_4$ in Section~\ref{sec:F4}, and $E_6$ in Section~\ref{sec:E6}. These exceptional Lie groups have found various applications in high energy physics in the context of gauge theories
and model building~\cite{Gursey:1975ki, Holland:2003jy, Shahlaei:2018lth, ramond2003exceptional, Deldar:2011fh, HosseiniNejad:2014giv, Deppisch:2016xlp,
Todorov:2018yvi,Todorov:2018mwd,Corradetti:2021abn}. Given the significant mathematical complexity of these groups, our ability to successfully derive their algebras attests to the robustness and generality of our ML techniques. 

\begin{figure}[t]    
    \centering
    \includegraphics[width=0.9\columnwidth]{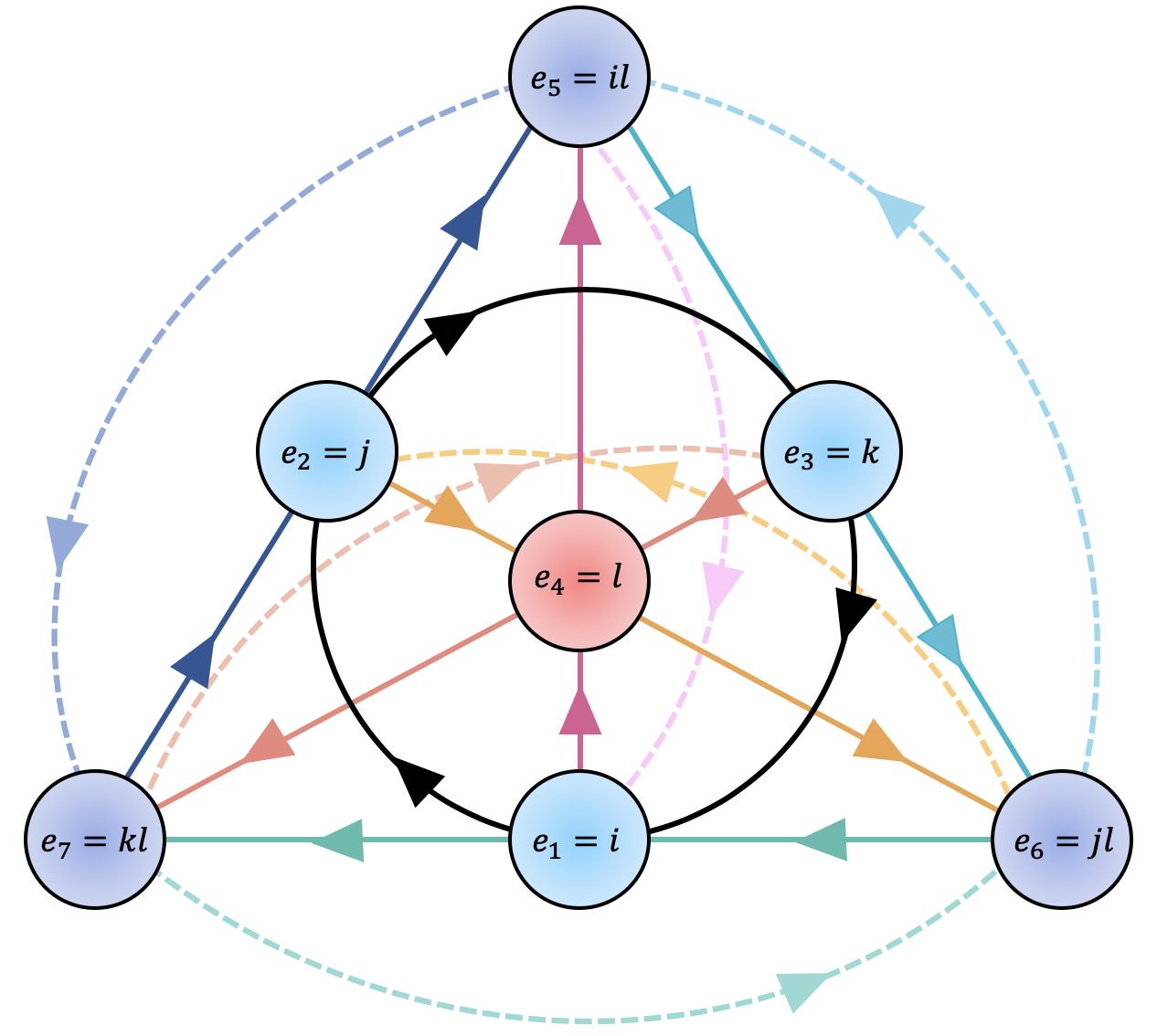} 
    \caption{Fano plane illustrating the multiplication rules for the imaginary unit octonions $e_1, \ldots, e_7$ in our conventions. For each triple $e_i$, $e_j$ and $e_k$ connected by a solid line, the result of the multiplication $e_i e_j$ is equal to $+e_k$ ($-e_k$) when going along (against) the arrows. The dashed lines have been added to guide the eye in following each triplet cycle. }
    \label{fig:octonion_multiplication}
\end{figure}

The smallest exceptional Lie group is the $G_2$ group, which has rank $2$ and dimension $14$. This group emerges as the automorphism group of the octonion algebra~\cite{Baez:2001dm}. An octonion $\mathbf o$ is a linear combination
\beq
\mathbf o = \sum_{i=0}^7 x^{(i)} e_i
\label{eq:odefinition}
\eeq
of the unit octonions $e_i$ with real coefficients $x^{(i)}$. Here $e_0$ is the real element which obeys $e_0^2=+1$ and can be identified with the real number 1. The remaining $e_1, \ldots, e_7$ are the seven imaginary unit octonions which obey $e_i^2=-1$. Their multiplication rules can be visualized in the Fano plane of Figure~\ref{fig:octonion_multiplication}. The figure shows only one of 480 possible definitions for octonion multiplication with $e_0 = 1$; the other definitions are isomorphic and can be obtained by permuting and/or changing the signs of the imaginary basis elements. 

The group $G_2$  has a fundamental representation of dimension 7, hence in this section $\mathbf x \in \mathbb R^{7}$. The components of the feature vector $\mathbf x$ will be identified with the coefficients $x^{(i)}$, $i=1,\ldots,7$, of the {\em imaginary} unit octonions in (\ref{eq:odefinition}). Correspondingly, the generators $\{J\}$ will be real $7\times7$ matrices.

In general, the exceptional groups preserve $K$ vector oracles $\varphi^{(1)}, \ldots, \varphi^{(K)}$, where each component $\varphi^{(i)}$ represents an invariant polynomial characteristic of the group. The group $G_2$ preserves $K=2$ such polynomials. The first one is the norm of a purely imaginary octonion,
\beq 
\varphi_{G_2}^{(1)} (\mathbf{x})= \sum_{i=1}^7 \bigl(x^{(i)}\bigr)^2.
\label{eq:oracleG2_1}
\eeq
To define the second oracle, we need to introduce the real part of the product of three octonions $\mathbf o_1$, $\mathbf o_2$ and $\mathbf o_3$
\beq
\Re \bigl( \mathbf o_1 \mathbf o_2 \mathbf o_3 \bigr) = 
\sum_{i,j,k=0}^{7} \mathcal{D}_{ijk} \,x_1^{(i)} \, x_2^{(j)}\, x_3^{(k)}.
\label{eq:tripleproduct}
\eeq
\begin{figure}[t]
\centering
\includegraphics[height=0.14\textwidth]{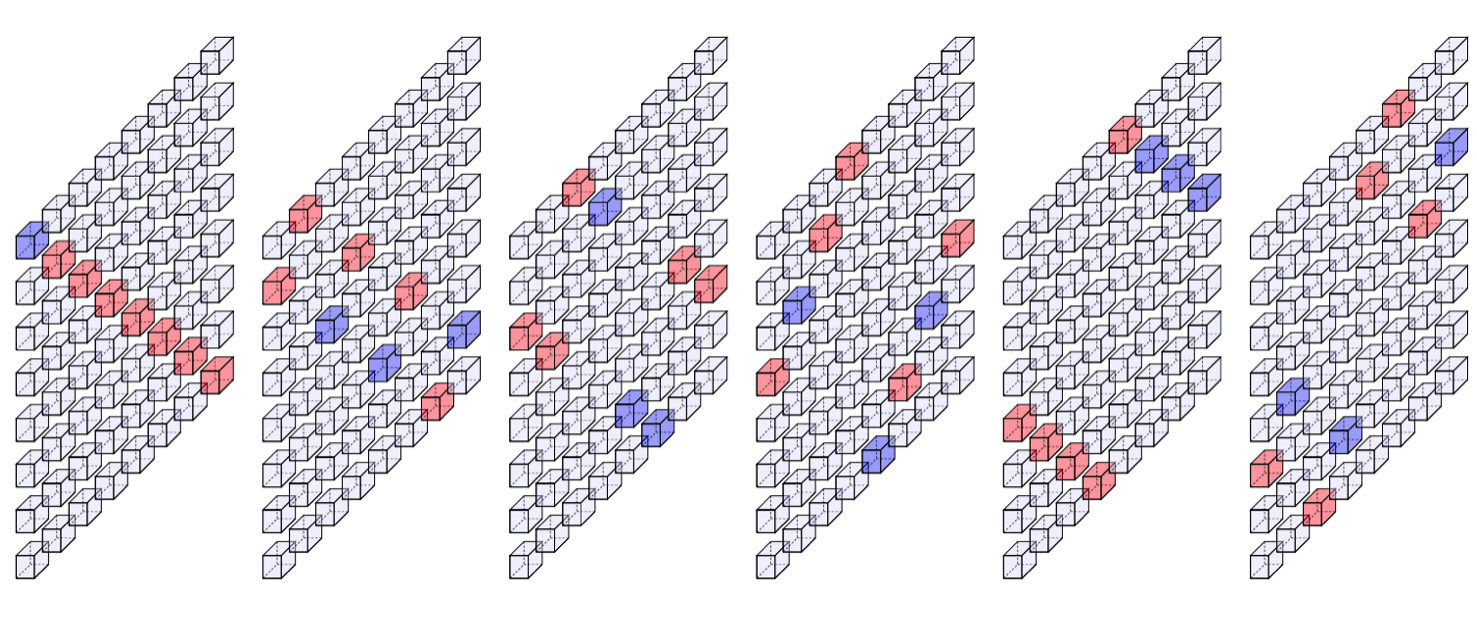} 
\includegraphics[height=0.14\textwidth]{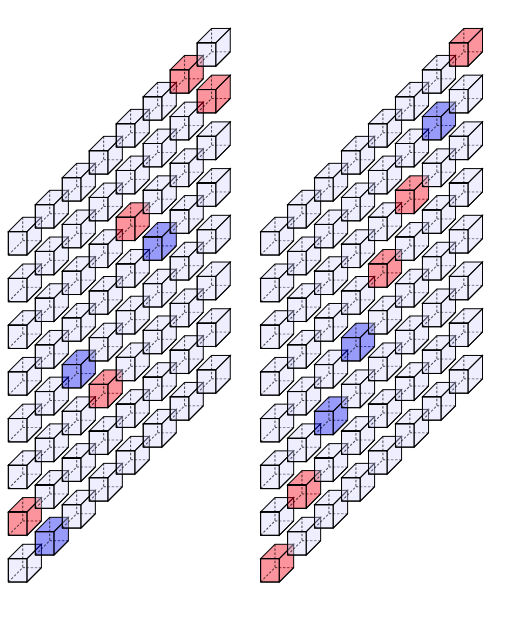} 
\caption{A pictorial visualization of the rank three tensor $\mathcal{D}_{ijk}$ defined in Eq.~(\ref{eq:tripleproduct}). Each plane represents a slice at a fixed $i=0,1,\ldots,7$ (from left to right). The blue, red and grey boxes indicate coefficient values of $+1$, $-1$ and $0$, respectively.  
}
\label{G2:octonion_cross}
\end{figure}
\begin{figure*}[t]
    \centering 
    \includegraphics[width=0.98\textwidth]{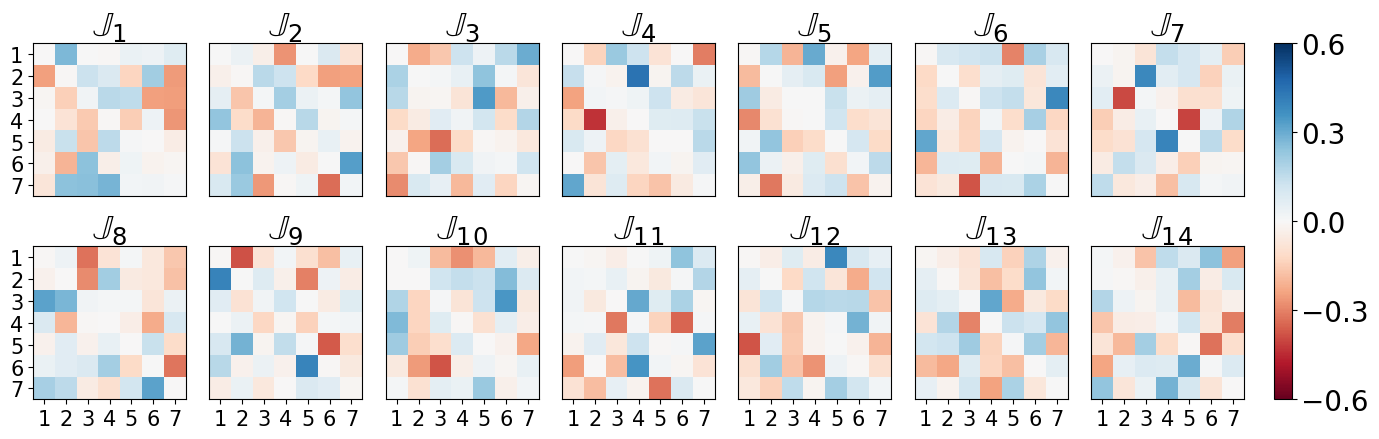} \\
    \vspace{2mm}
    \includegraphics[width=0.98\textwidth]{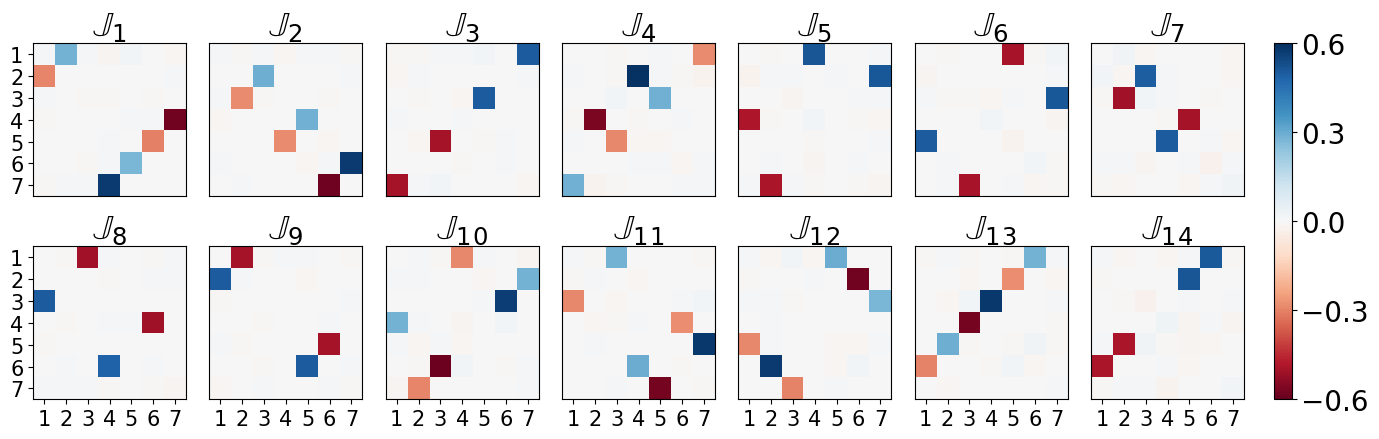} 
    \caption{The fourteen $G_2$ generators learned with the greedy method (top panels) and the result from their sparsification (bottom panels).  In this and all subsequent such figures, each panel represents a learned generator ${\mathbb J}_\alpha$ in matrix form, where the values of the individual elements of the matrix are indicated by the color bar. 
    }
    \label{G2:generators}
\end{figure*}
Using the multiplication rules from Figure~\ref{fig:octonion_multiplication}, the components of the rank three tensor $\mathcal{D}_{ijk}$ can be easily derived and are shown in Figure~\ref{G2:octonion_cross}. The group $G_2$ preserves the real component of the product of three {\em purely imaginary} octonions, hence the second $G_2$ oracle is (note that the sums start from 1 instead of 0)
\beq
\varphi^{(2)}_{G_2}({\bf x}_1, {\bf x}_2, {\bf x}_3) =
\sum_{i,j,k=1}^{7} \mathcal{D}_{ijk}
\,x_1^{(i)} \, x_2^{(j)}\, x_3^{(k)}.
    \label{eq:oracleG2_2}
\eeq

For the training of the $G_2$ generators, we use $m=900$ samples (7-dimensional vectors in $\mathbb R^7$) and fix $\varepsilon = 10^{-3}$. For the computation of the second oracle (\ref{eq:oracleG2_2}) we split the sample into three equally sized groups of $300$, from which we draw the three vectors $\mathbf x_1$, $\mathbf x_2$ and $\mathbf x_3$. The training  was done with the {\sc Adam} optimizer~\cite{kingma2017adam} for 1,000 epochs and with learning rate of $2.5\times 10^{-2}$. 

Our results for the learned $G_2$ generators are shown in Figure~\ref{G2:generators}. The panels in the top two rows depict the 14 generators found by the greedy algorithm (as expected, the algorithm failed to find a valid 15th generator, where even after 10,000 epochs, the loss stayed of order 1). The panels in the bottom two rows show the corresponding results after applying the sparsification procedure of Section~\ref{sec:sparsification}. Note that all found generators are antisymmetric ($G_2$ is a subgroup of $SO(7)$), and that they can be organised into 7 pairs which share common matrix elements:
$\mathbb J_1$ and $\mathbb J_9$,
$\mathbb J_2$ and $\mathbb J_7$,
$\mathbb J_3$ and $\mathbb J_4$,
$\mathbb J_5$ and $\mathbb J_{10}$,
$\mathbb J_6$ and $\mathbb J_{12}$,
$\mathbb J_8$ and $\mathbb J_{11}$,
$\mathbb J_{13}$ and $\mathbb J_{14}$. 
Note that in each pair, one generator has six non-zero elements, while the other has only four. Therefore our version is sparser than the conventional textbook representation, in which each generator has six nonvanishing elements.

\section{The Exceptional Group \texorpdfstring{$F_4$}{F4}}
\label{sec:F4}

In this and the next section we follow the notation of refs.~\cite{Bernardoni:2007rf} and \cite{Bernardoni:2007rd}, where the generators for $F_4$ and $E_6$ have been explicitly derived. The exceptional group $F_4$ has rank $4$ and dimension 52. It is the automorphism group of the Jordan algebra 
\beq
{\mathfrak h}_3 =
\left(
\begin{matrix}
r_1 & {\mathbf o}_1 & {\mathbf o}_2 \\
{\mathbf o}^\ast_1 & r_2 & {\mathbf o}_3 \\
{\mathbf o}^\ast_2 & {\mathbf o}^\ast_3 & r_3
\end{matrix}
\right),
\label{eq:Jordandef}
\eeq
where $r_a$ are three real numbers ($a=1,2,3$), and $\mathbf{o}_a$ are three octonions \cite{Bernardoni:2007rf}. The asterisk notation in (\ref{eq:Jordandef}) stands for octonion conjugation
\beq
\mathbf o^\ast = x^{(0)} e_0-\sum_{i=1}^7 x^{(i)} e_i.
\label{eq:oconjugatedefinition}
\eeq
The fundamental representation is of dimension 26. Following \cite{Bernardoni:2007rf,Bernardoni:2007rd}, we find it convenient to work in $\mathbb R^{27}$ instead and map the components of the feature vector $\mathbf x$ onto $r_a$ and $\mathbf{o}_a$ as follows
\begin{subequations}
\begin{alignat}{2}
r_1 &= x^{(1)}, \qquad &\mathbf{o}_1 &= \sum_{i=0}^7 x^{(2+i)}\, e_i,  \\
r_2 &= x^{(18)},       &\mathbf{o}_2 &= \sum_{i=0}^7 x^{(10+i)}\, e_i,   \\
r_3 &= x^{(27)},       &\mathbf{o}_3 &= \sum_{i=0}^7 x^{(19+i)}\, e_i.
\end{alignat}
\label{eq:F4octonions}
\end{subequations}

The group $F_4$ preserves $K=3$ different oracles (invariant polynomials), which can be expressed in terms of the variables (\ref{eq:F4octonions}) as follows:
\begin{align}
\varphi^{(1)}_{F_4}(\mathbf x) &= \text{Tr} \,\mathfrak h_3 = \sum_{a=1}^3 r_a = x^{(1)} + x^{(18)} + x^{(27)} , \label{F4oracle1}
\end{align}
\begin{align}
\varphi^{(2)}_{F_4}(\mathbf x) &= \text{Tr} \,\mathfrak h^2_3 =\sum_{a=1}^3 \left(\, r_a^2 + 2\, 
|{\mathbf o}_a|^2\,\right)  \nonumber \\[2mm] 
&=2 \sum_{i=1}^{27} \left( x^{(i)}\right)^2 - \sum_{i\in \{1,18,27\}} \left( x^{(i)}\right)^2 ,
\label{F4oracle2}
\end{align}
\begin{align}
\varphi^{(3)}_{F_4}(\mathbf x) &= \det \mathfrak h_3 =
r_1r_2r_3 - \sum_{a=1}^3 r_a\, |{\mathbf o}_{4-a}|^2 + 2 \Re \left( {\mathbf o}_3 {\mathbf o}_2^\ast {\mathbf o}_1  \right) \nonumber \\[2mm] 
&=x^{(1)} x^{(18)} x^{(27)} - x^{(1)} \sum_{i=19}^{26} \left( x^{(i)}\right)^2 \nonumber \\
& - x^{(18)} \sum_{i=10}^{17} \left( x^{(i)}\right)^2 
  - x^{(27)} \sum_{i=2}^{9} \left( x^{(i)}\right)^2 \nonumber \\
&+ 2\, \sum_{i,j,k=0}^{7} \mathcal{D}_{ijk}\, x^{(19+i)} \,x^{(10+j)} \, x^{(2+k)} \, (2\delta_{j,10}-1),
\label{F4oracle3}
\end{align} 
where the tensor $\mathcal{D}_{ijk}$ was defined in Eq.~(\ref{eq:tripleproduct}) and pictorially illustrated in Figure~\ref{G2:octonion_cross}. The additional factor of $(2\delta_{j,10}-1)$ in the last line flips the sign of the $x^{(11)}, x^{(12)}, \ldots, x^{(17)}$ factors in the sum and thus accounts for the conjugation of ${\mathbf o}_2$ in the triple product ${\mathbf o}_1 {\mathbf o}_2^\ast {\mathbf o}_3$.

\begin{figure*}[tbhp]
\centering
\includegraphics[width=0.95\linewidth]{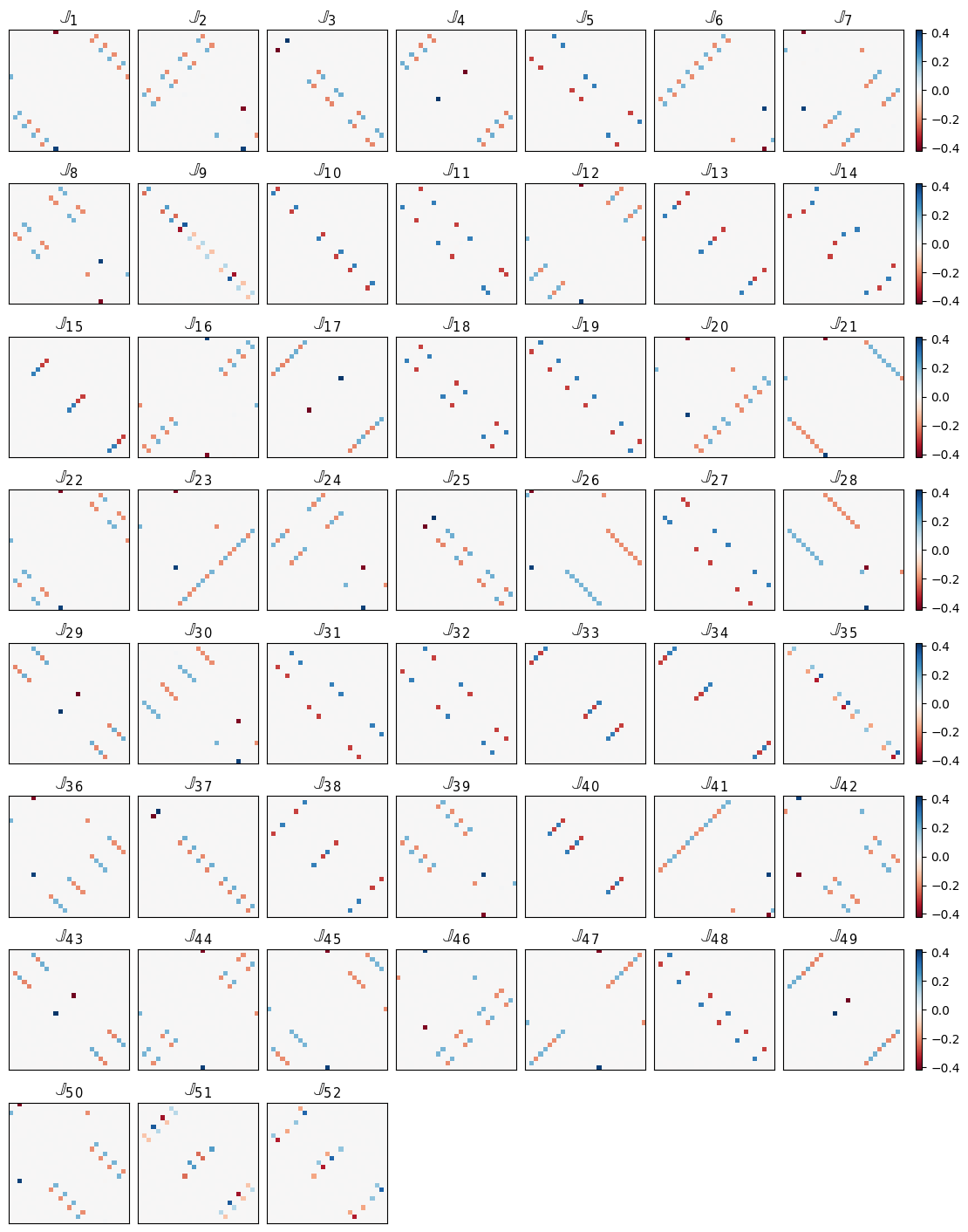}    \caption{The learned $27\times 27$ sparse generators $J_\alpha$, $\alpha=1,\ldots,52$, for the case of $F_4$. }
\label{F4:generators}
\end{figure*}

The results from the training of 52 $F_4$ generators with the greedy algorithm and with the oracles (\ref{F4oracle1}-\ref{F4oracle3}) are shown in Figure~\ref{F4:generators}. We used $m=900$ samples of the $n=27$ input features $x^{(i)}$, organized as in (\ref{eq:F4octonions}). The training ran over up to 26,000 epochs with the {\sc Adam} optimizer, $\varepsilon = 10^{-4}$ and learning rate of $5\times 10^{-4}$. A 53rd generator was not found after more than 250,000 epochs.

The numerically derived generators in Figure~\ref{F4:generators} can be compared against explicit analytical constructions of the $F_4$ generators in the literature. For example, we have verified that the set of generators in Figure~\ref{F4:generators} is isomorphic (in the sense that the matrix $O$ relating the two sets as in (\ref{eq:sparse_rotation}) is orthogonal) to the set of fifty two $27\times 27$ matrices $\mathbb C_\alpha$ listed in Appendix C of \cite{Bernardoni:2007rf}. This comparison is rather nontrivial in light of the following differences between the two studies:
\begin{itemize}
\item {\em Octonion multiplication table.} The multiplication rules satisfied by the imaginary unit octonions $e'_i$ in \cite{Bernardoni:2007rf} are different from those of Figure~\ref{fig:octonion_multiplication}. The two bases are related as $e'_1 = e_1$, $e'_2 = e_2$, $e'_3 = e_4$, $e'_4 = e_3$, $e'_5 = e_6$, $e'_6 = -e_7$, $e'_7 = e_5$. As a result, the two sets of generators appear visually different, as the non-vanishing components of the respective matrices are located in different rows and/or columns. As it turns out, the conventions of Figure~\ref{fig:octonion_multiplication} happen to produce an aesthetically more pleasing result --- note how all non-vanishing entries in Figure~\ref{F4:generators} are nicely lined up diagonally, while the corresponding patterns in Appendix C of \cite{Bernardoni:2007rf} appear more scattered and chaotic.
\item {\em Normalization.} The $F_4$ generators $\mathbb C_\alpha$ in \cite{Bernardoni:2007rf} are normalized as $\text{Tr}(\mathbb C_\alpha \mathbb C_\beta)=-\,6\,\delta_{\alpha\beta}$, while those in Figure~\ref{F4:generators} are normalized as $\text{Tr}(\mathbb J_\alpha \mathbb J^T_\beta)=\delta_{\alpha\beta}$ as per (\ref{eq:LossNormalization}) and (\ref{eq:LossOrthogonality}).
\item {\em Basis of $\mathbb R^{27}$.} Since the first oracle (\ref{F4oracle1}) is linear in $\mathbf x$, one can apply an orthogonal rotation to $r_1$, $r_2$ and $r_3$, so that one of the coordinates in the new basis, say the last one, is directly proportional to $\varphi_{F_4}^{(1)}$ \cite{Bernardoni:2007rf}. This choice has the advantage that the last row and the last column of each generator matrix are filled with zeros, as in \cite{Bernardoni:2007rf}, confirming that the fundamental representation of $F_4$ is 26-dimensional. If we apply the same rotation to the learned generators in Figure~\ref{F4:generators}, the resulting matrices all have zeros in their last rows and columns as well.
\end{itemize}

\section{The Exceptional Group \texorpdfstring{$E_6$}{E6}}
\label{sec:E6}

\begin{figure*}[t]
\centering
\includegraphics[width=0.95\linewidth]{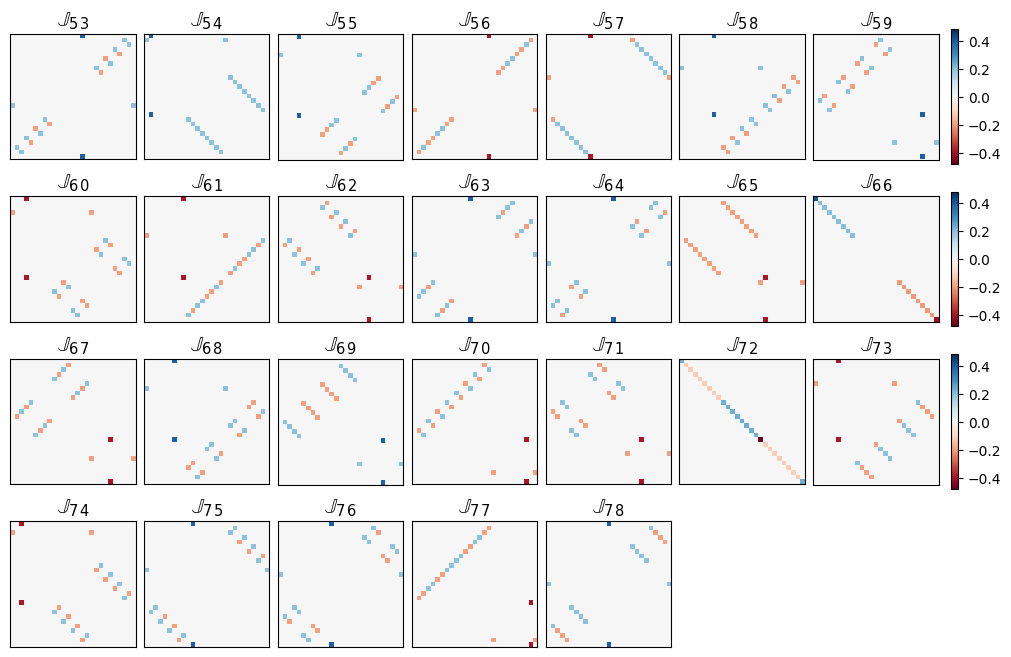}
\caption{The additional learned $27\times 27$ sparse generators $J_\alpha$, $\alpha=53,\ldots,78$, for the case of $E_6$. The hyperparameters used in the training of the single non-sparse generator $\mathbb J_{53}$ used to seed the algorithm were as in Section~\ref{sec:F4}. }
\label{E6:generators}
\end{figure*}

The exceptional group $E_6$ is of rank $6$ and dimension $78$. Following \cite{Bernardoni:2007rd}, we work in the real case, which results in the split form of the $E_6$ algebra, with signature $(52,26)$. (By multiplying the 26 added generators by $i$, the algebra remains real, and the Killing form becomes the compact one. For details, we refer the interested reader to \cite{Bernardoni:2007rd}.)  The relevant $27$ variables are those in (\ref{eq:F4octonions}). In this case, we require invariance with respect to only the last oracle, (\ref{F4oracle3}), but not the first two, (\ref{F4oracle1}) and (\ref{F4oracle2}). This allows for the presence of $78-52=26$ additional generators beyond those of Fig.~\ref{F4:generators}. Our goal in this section will be to derive those additional 26 generators which are not contained in the $F_4$ subalgebra of $E_6$. 

The straightforward approach to deriving all $E_6$ generators would be to apply the greedy algorithm and learn from scratch 78 generators preserving the oracle (\ref{F4oracle3}). However, we can save a significant amount of work by leveraging the learned $F_4$ generators from the previous section which already satisfy (\ref{F4oracle3}) by definition. In other words, given the 52 already discovered $F_4$ generators, we are looking to find the additional 26 which complete $F_4$ to $E_6$. This is the perfect setup for applying the Lie bracket trick --- treat the $F_4$ generators as the starter set $\{\mathbb J_1, \ldots, \mathbb J_{52}\}$, learn {\em a single} new non-sparse generator $\mathbb J_{53}$ with the greedy algorithm, and then hand those $52+1$ generators over to the LBT algorithm and let it do its job. The result from this procedure (after the corresponding sparsification of the 26 new generators among themselves) is shown in Figure~\ref{E6:generators}. Notably, the LBT algorithm was able to find {\em all} of the missing 25 generators from a single non-sparse seed $\mathbb J_{53}$. The key to this was that the seed $\mathbb J_{53}$ was non-sparse, and therefore a linear combination involving a large number of the canonical sparse $E_6$ generators.

The meachine-learned generators shown in Figure~\ref{E6:generators} can be verified against analytically derived results in the literature. We have compared our results to the additional 26 $E_6$ generators $\mathbb C_{53}, \ldots, \mathbb C_{78}$ listed in Appendix C of \cite{Bernardoni:2007rd} and found perfect agreement, once we account for the differences in our conventions (see discussion at the end of Section~\ref{sec:F4}). For example, the LBT algorithm correctly finds exactly two diagonal generators, namely $\mathbb J_{66}$ and $\mathbb J_{72}$ (respectively proportional to $\mathbb C_{53}$ and $\mathbb C_{70}$ in \cite{Bernardoni:2007rd}). This is expected, since the rank of $E_6$ is larger than the rank of $F_4$ by 2. Also note that all generators are antisymmetric in the octonionic indices $2-17, 19-26$. Once again, we find regular linear patterns in the locations of the nonzero matrix elements in the generators in Figure~\ref{E6:generators}. 

\section{Summary}
\label{sec:summary}

Discovering symmetries in data is crucial for both fundamental theory and data science. Recent advancements in ML algorithms, coupled with the growth in computational resources, have facilitated progress in this area. However, when dealing with highly complex and multidimensional problems, training machine-learning models remains a significant bottleneck. This paper introduces two novel algorithms that greatly accelerate the symmetry discovery process. The new methods were rigorously tested and showcased using examples of symmetries from the exceptional groups $G_2$, $F_4$, and $E_6$. Remarkably, the symmetry generators were learned accurately and in a nicer form than the conventional representations found in textbooks. With these groundbreaking techniques at our disposal, we can now confidently tackle even more intricate and challenging problems in mathematical physics, particle phenomenology, and data analysis.

{\bf Acknowledgements.}
We are greatly indebted to P.~Ramond for setting us on this exceptional path by suggesting this project.  
We thank S.~Gleyzer, R.~Houtz, K.~Kong, S.~Mrenna, H.~Prosper and P.~Shyamsundar for useful discussions. This work is supported in part by US~Department of Energy award DE-SC0022148.

\appendix
\section{Loss function}
\label{app:loss}

In this appendix we adapt the loss functions from \cite{Forestano:2023fpj,Forestano:2023qcy} to the case where we train a single generator $\mathbb G$. 
The loss function is chosen to ensure that $\mathbb G$ has the following properties:

{\bf Invariance:} preserving the values of a vector oracle $\vec{\varphi}(\mathbf{x})$ for all sampled datapoints $\left\{\mathbf{x}\right\}$ under a given candidate transformation $\mathbb G$:
\beq
    L_\text{inv}\bigl(\mathbb G, \{\mathbf x\}\bigl) = \frac{1}{m\varepsilon^2}\sum_{i=1}^m \Bigl[ \vec\varphi\bigl(\mathbf{x}_i + \varepsilon {\mathbb G}\cdot\mathbf{x}_i\bigr)-\vec\varphi(\mathbf{x}_i) \Bigr]^2 \, ,
\label{eq:LossInvariance}    
\eeq
where ``$\cdot$" denotes ordinary tensor multiplication and the arrow vector notation implies summation over the $K$ oracle components.

{\bf Normalization:} ensuring that the transformation $\mathbb G$ is not trivial and normalized to 1:
\beq
L_\text{norm}\bigl(\mathbb G\bigr) =  \Bigl[ \text{Tr}\Bigl({\mathbb G}\cdot {\mathbb G}^T\Bigr) - 1\Bigr]^2 .
\label{eq:LossNormalization}    
\eeq

{\bf Orthogonality.} This condition is used if we already have some existing orthonormalized generators $\{J\}$ and want to find an additional generator $\mathbb G$. It guarantees that the transformation $\mathbb G$  is not in the set $\{J\}$: 
\beq
 L_\text{ortho}\bigl(\mathbb G, \{\mathbb J\} \bigr) 
 = 
 \sum_{\substack{\alpha }} \Bigl[ \text{Tr}\bigl({\mathbb G}\cdot {\mathbb J}_\alpha^T\bigr)\Bigr]^2.
\label{eq:LossOrthogonality}
\eeq

{\bf Sparsity:} an additional loss term designed to encourage sparsity was introduced in Ref.~\cite{Forestano:2023qcy} :
\beq
L_\text{sp} \bigl(\mathbb G \bigr) =
\sum_{j,k=1}^n
\sum_{\substack{j',k'=1}}^n
\left|{\mathbb G}^{(jk)}{\mathbb G}^{(j'k')}\right| \Bigl(1-\delta_{jj'}\delta_{kk'}\Bigr),
\label{eq:LossSparsity}
\eeq
where ${\mathbb G}^{(jk)}$ denotes the $jk$-component of $\mathbb G$.

In this study, the total loss function was formed as a plain sum of the relevant individual terms (\ref{eq:LossInvariance}-\ref{eq:LossSparsity}).

\bibliographystyle{elsarticle-num} 
\bibliography{references}

\end{document}